\documentclass[aps,prd,floatfix,nofootinbib,showpacs,twocolumn,superscriptaddress]{revtex4-1}

\usepackage{mathrsfs}
\usepackage{amsmath}
\usepackage{amssymb}
\usepackage{bbold}
\usepackage{color}
\usepackage{graphicx}
\usepackage{soul}

\usepackage[mathscr,scaled=1.15]{urwchancal}
\DeclareFontFamily{OT1}{pzc}{}
\DeclareFontShape{OT1}{pzc}{m}{it}%
{<-> s * [1.15] pzcmi7t}{}
\DeclareMathAlphabet{\mathpzc}{OT1}{pzc}{m}{it}

\definecolor{purple}{rgb}{0.5,0,0.5}
\definecolor{blue}{rgb}{0.0,0,0.9}

\begin{document}

\title{Leading-twist distribution amplitudes of scalar- and vector-mesons}

%\author{Bo-Lin Li}
\author{B.-L. Li}
\email{libolin0626@126.com}
\affiliation{Department of Physics, Nanjing University, Nanjing, Jiangsu 210093, China}

\author{L. Chang}
\email{leichang@nankai.edu.cn}
\affiliation{School of Physics, Nankai University, Tianjin 300071, China}

\author{M. Ding}
\email{mhding@nankai.edu.cn}
\affiliation{School of Physics, Nankai University, Tianjin 300071, China}

\author{C.\,D.~Roberts}
\email{cdroberts@anl.gov}
\affiliation{Physics Division, Argonne National Laboratory, Argonne
Illinois 60439, USA}

\author{H.-S. Zong}
\email{zonghs@nju.edu.cn}
\affiliation{Department of Physics, Nanjing University, Nanjing, Jiangsu 210093, China}
\affiliation{State Key Laboratory of Theoretical Physics, Institute of Theoretical Physics, CAS, Beijing, 100190, China}

\date{16 August 2016}
%\date{15 July 2016}

\begin{abstract}
A symmetry-preserving truncation of the two-body light-quark bound-state problem in relativistic quantum field theory is used to calculate the leading-twist parton distribution amplitudes (PDAs) of scalar systems, both ground-state and radial excitations, and the radial excitations of vector mesons.  Owing to the fact that the scale-independent leptonic decay constant of a scalar meson constituted from equal-mass valence-constituents vanishes, it is found that the PDA of a given scalar system possesses one more zero than that of an analogous vector meson.  Consequently, whereas the mean light-front relative momentum of the valence-constituents within a vector meson is zero, that within a scalar meson is large, an outcome which hints at a greater role for light-front angular momentum in systems classified as $P$-wave in quantum mechanical models.  Values for the scale-dependent decay constants of ground-state scalar and vector systems are a by-product of this analysis, and they turn out to be roughly equal, \emph{viz}.\ $\simeq 0.2\,$GeV at an hadronic scale. In addition, it is confirmed that the dilation characterising ground-state PDAs is manifest in the PDAs of radial excitations too.  The impact of $SU(3)$-flavour symmetry breaking is also considered.  When compared with pseudoscalar states, it is a little stronger in scalar systems, but the size is nevertheless determined by the flavour-dependence of dynamical chiral symmetry breaking and the PDAs are still skewed toward the heavier valence-quark in asymmetric systems.
\end{abstract}

%%\keywords{
%%Abelian anomaly \and
%%continuum QCD \and
%%dynamical chiral symmetry breaking \and
%%$\pi$-meson elastic and transition form factors \and
%%parton distribution amplitudes}


\maketitle

\section{Introduction}
Scalar and vector mesons are intimately connected.  The vector Ward-Green-Takahashi (WGT) identity, typically associated with electromagnetic current conservation, ensures that, in the presence of a nonzero difference between the current-masses of a given channel's two valence-quarks $(m_{q_1} \neq m_{\bar q_2})$, a scalar vertex, including all its associated poles (scalar mesons), is indistinguishable from the longitudinal projection of the related vector vertex.  Consequently, notwithstanding the potential complications \cite{Agashe:2014kda, Pelaez:2015qba}, even light scalar mesons should contain a measurable component with standard mesonic character, \emph{i.e}.\ generated by a leading-twist quark-antiquark interpolating field \cite{Holl:2005st, Eichmann:2015cra}.

In the rest frame, the quark-antiquark component of light-quark scalar mesons is predominantly $S$-wave in character \cite{Maris:2000ig, Krassnigg:2009zh}.  Although this conclusion conflicts with notions derived from quantum mechanical two-body models, which describe scalar mesons as $^3P_0$ states, it should not be surprising because: the class of contact-interaction theories generate a $\sigma$-meson as the chiral partner of the pion \cite{Vogl:1991qt, Klevansky:1992qe}, also an $S$-wave state; and a vector$\otimes$vector contact-interaction produces Bethe-Salpeter amplitudes that are independent of $q_1$-$\bar q_2$ relative momentum \cite{Roberts:2011cf}, in which case orbital angular momentum within the bound-state is greatly suppressed.  On the other hand, a $P$-wave component in the scalar meson grows with increasing current-quark mass, so that the quark-model description of scalar mesons is valid for $m_{q_1}, m_{\bar q_2} \gg \Lambda_{\rm QCD}$.   Analogously, light-quark vector-mesons contain both $S$- and $D$-wave components of commensurate size, but the $D$-wave component diminishes with increasing current-quark mass, so that these states may be described as $^3S_1$ systems in the heavy-heavy limit \cite{Gao:2014bca}.  Consequently, a given scalar meson may be understood as a $\Delta L =1 $ orbital excitation of a related vector meson when $m_{q_1}, m_{\bar q_2} \gg \Lambda_{\rm QCD}$.

There is at least one significant difference between scalar and vector mesons, however.  Namely, when $m_{q_1} = m_{\bar q_2}$, vector mesons possess two nonzero decay constants, one of which describes the bound-state's leptonic decay, but the analogous scalar-meson decay constant is identically zero: $f_\sigma\equiv 0$.  This follows from the WGT identity and is simply the statement that a $J^{PC}=1^{--}$ current cannot connect a $0^{++}$ bound-state to the $0^{++}$ vacuum.  These observations translate into the result that whilst the two leading-twist vector-meson parton distribution amplitudes (PDAs) possess a nonzero leading Mellin moment at any finite renormalisation scale, this moment vanishes for all scalar mesons whose valence degrees-of-freedom satisfy $m_{q_1} = m_{\bar q_2}$, irrespective of the size of the common current-mass, large or small.  This feature must entail considerable differences between the leading-twist PDAs of scalar and vector mesons.

Amongst other things, leading-twist meson PDAs play an important role in the analysis and understanding of hard exclusive processes, such as pseudoscalar meson electromagnetic form factors \cite{Lepage:1979zb, Efremov:1979qk, Lepage:1980fj, Horn:2016rip} and diffractive vector-meson production \cite{Gao:2014bca, Forshaw:2010py, Forshaw:2012im}, and also in the study of $CP$-violation via nonleptonic decays of heavy-light mesons \cite{Neubert:1998jq, Beneke:2001ev, Cheng:2005nb, Cheng:2009xz, ElBennich:2009da, Shi:2015esa}.  Additionally, they provide deep insights into the structure of hadron bound-states, revealing, \emph{e.g}.\ how mass is distributed \cite{Roberts:2016vyn} and momentum is shared amongst an hadron's constituents.  Substantial value is therefore attached to their computation in frameworks with a traceable connection to QCD.

Using QCD's Dyson-Schwinger equations (DSEs) \cite{Bashir:2012fs, Cloet:2013jya, Roberts:2015lja, Pennington:2016dpj}, low-twist PDAs of pseudoscalar- and vector-meson ground-states and pseudoscalar-meson radial-excitations have recently been computed \cite{Gao:2014bca, Shi:2015esa, Ding:2015rkn, Li:2016dzv}; and, where a meaningful comparison is possible, the results agree with those determined via numerical simulations of lattice-regularised QCD \cite{Cloet:2013tta, Segovia:2013eca, Horn:2016rip}.  Hence, given the features highlighted above, we consider it interesting to use this approach to compute the leading-twist PDAs of scalar-meson ground-states and radial excitations, and the radial excitations of vector mesons, all constituted from light quarks.  The complete body of results thus obtained should prove useful in both developing novel insights into hadron structure and constraining phenomenological applications of hard scattering formulae in a wide variety of processes.

We provide the background for our calculations in Sec.\,\ref{SecScalars}, including details of the gap and Bethe-Salpeter interaction kernels, and a description of the manner by which we recover PDAs from their Mellin moments.  This leads naturally to the presentation and discussion of results for the PDAs of a range of $0^+$ quark-antiquark systems.  The same methods are employed in Sec.\,\ref{SecVectors} to analyse the PDAs of radially-excited light-quark vector mesons.  We summarise and outline some future prospects in Sec.\,\ref{epilogue}.

\section{Scalar mesons}
\label{SecScalars}
\subsection{Prelude}
All scalar mesons that possess nonzero overlap with the interpolating field $\bar q_1 \mathbf{I}_D q_2$, where $\mathbf{I}_D$ is a $4\times 4$ identity matrix acting on spinor indices, appear as poles in the Bethe-Salpeter equation whose inhomogeneity is $(1/2)\lambda^{q_1 q_2}\mathbf{I}_D$, where $\lambda^{q_1 q_2}$ is a matrix specifying the flavour structure of the system.  Denoting the total momentum flowing into the scalar vertex by $P$, a pole at $P^2 + s_M=0$ need not lie on the real-$P^2$ axis.\footnote{We use a Euclidean metric:  $\{\gamma_\mu,\gamma_\nu\} = 2\delta_{\mu\nu}$; $\gamma_\mu^\dagger = \gamma_\mu$; $\gamma_5= \gamma_4\gamma_1\gamma_2\gamma_3$, tr$[\gamma_5\gamma_\mu\gamma_\nu\gamma_\rho\gamma_\sigma]=-4 \epsilon_{\mu\nu\rho\sigma}$; $\sigma_{\mu\nu}=(i/2)[\gamma_\mu,\gamma_\nu]$; $a \cdot b = \sum_{i=1}^4 a_i b_i$; and $P_\mu$ timelike $\Rightarrow$ $P^2<0$.}  Its real and imaginary parts provide information about the mass and width of the state: $\surd s_M = m_M - i \Gamma_M/2$.  The residue at the vertex pole is the scalar meson's Bethe-Salpeter amplitude, which has the following form owing to the requirements of Poincar\'e covariance:
\begin{eqnarray}
\nonumber
\lefteqn{\Gamma_{\sigma}(\ell;P) =  \lambda^{q_1 q_2} \mathbf{I}_D
\big[ E_{\sigma}(\ell;P) + i \gamma\cdot P F_{\sigma}(\ell;P)   }\\
&&  \quad\quad  +i \gamma\cdot \ell \, G_{\sigma}(\ell;P) + \sigma_{\mu\nu} \ell_\mu P_\nu H_{\sigma}(\ell;P) \big], %\rule{0.7em}{0ex}
\label{BSS}
\end{eqnarray}
where $\ell = (l_{q_1} + l_{\bar q_2})/2$, with $l_{q_1} = l + \eta P$, $l_{\bar q_2}=l-(1-\eta) P$, being the momenta attached to the quark and antiquark legs, respectively.  Owing to Poincar\'e covariance, no observable depends on $\eta\in[0,1]$, \emph{i.e}.\ the definition of the relative momentum.

Attaching propagator legs to the amplitude, one obtains the Bethe-Salpeter wave function:
\begin{equation}
\chi_\sigma(\ell;P) = S_{1}(l_{q_1}) \Gamma_\sigma(\ell;P) S_{2}(l_{\bar q_2})\,,
\label{chiS}
\end{equation}
where $S_{1,2}$ are propagators associated with quark flavours $q_{1,2}$, respectively.  This wave function can be expressed in a form analogous to Eq.\,\eqref{BSS}, using scalar functions $\chi_\sigma^{E,F,G,H}$, in which case $\chi_\sigma^{E,F}$ are associated with $L=0$ and $\chi_\sigma^{G,H}$ with $L=1$ in the meson's rest frame \cite{LlewellynSmith:1969az, Krassnigg:2009zh}.

The two simplest projections of the wave function onto the origin in configuration space are
\begin{subequations}
\begin{align}
\label{fsigma}
f_\sigma P_\mu &= {\rm tr} \,Z_2 \! \int_{dl}^\Lambda \tfrac{1}{2} \lambda^{q_1 q_2} \gamma_\mu \chi_\sigma(\ell;P) \,,\\
m_\sigma \tilde f_\sigma(\zeta) := \rho_\sigma(\zeta) & = {\rm tr} \, Z_4 \! \int_{dl}^\Lambda \tfrac{1}{2} \lambda^{q_1 q_2}  \mathbf{I}_D \chi_\sigma(\ell;P) \,,
\label{rhosigma}
\end{align}
\end{subequations}
where $\int_{dl}^\Lambda$ is shorthand for a Poincar\'e-invariant regularisation of the four-dimensional momentum integral, with $\Lambda$ the regularisation scale, and $Z_{2,4}$ are renormalisation constants for the quark wave function and scalar vertex, respectively.  The vector projection, Eq.\,\eqref{fsigma}, defines the scalar meson's leptonic decay constant, which vanishes for $J^{PC}=0^{++}$ states, whereas the scalar projection, Eq.\,\eqref{rhosigma}, is always nonzero, increasing with renormalisation scale, $\zeta$, just like the chiral condensate \cite{Brodsky:2012ku}.

The leading-twist PDA of a scalar meson is connected with Eq.\,\eqref{fsigma}, \emph{viz}.\
\begin{align}
\nonumber
\phi_\sigma(x) & = {\rm tr} \,Z_2 \int_{dl}^\Lambda \! \delta(n\cdot l_{q_1} - x \,n\cdot P) \\
    & \quad \times \tfrac{1}{2} \lambda^{q_1 q_2} \gamma\cdot n \chi_\sigma(\ell;P) \,,
\end{align}
where $n$ is a light-like four-vector, $n^2=0$.  Thus defined, the distribution has mass-dimension ``one'': given that for light-quarks the leptonic decay constant is either zero or small, it does not serve as a useful mass scale; and the distribution's Mellin moments can be obtained via
\begin{align}
\nonumber
\langle x^m\rangle (n\cdot P)^{m+1} & = {\rm tr} \,Z_2 \int_{dl}^\Lambda \! (n\cdot l_{q_1} )^m  \\
& \quad \times \tfrac{1}{2} \lambda^{q_1 q_2}\, \gamma\cdot n \chi_\sigma(\ell;P) \,.
\label{Mellinmom}
\end{align}

As demonstrated elsewhere \cite{Chang:2013pq}, one can learn what to expect for the pointwise behaviour of $\phi_\sigma(x)$ in a theory whose interaction is $(1/k^2)^\nu$ vector-boson exchange by computing the result using
\begin{eqnarray}
\label{pointS}
S(p) &=& [-i\gamma \cdot p + M] \Delta_M(p^2)\,, \\
\label{rhoznu}
\rho_\nu(z) &=& \frac{1}{\surd \pi}\frac{\Gamma(\nu + 3/2)}{\Gamma(\nu+1)}\,(1-z^2)^\nu\,,\\
\label{rhoEpi}
\Gamma_\sigma(l;P) & = &
\mathbf{I}_D \frac{M}{f_\pi} \!\! \int_{-1}^{1}\!\! \!dz \,\rho_\nu(z)
\tilde \Delta_M^\nu(l_{+z}^2)\,,
\end{eqnarray}
where $\Delta_M(s) = 1/[s+M^2]$, $\tilde \Delta_M = M^2 \Delta_M$, $l_{+z}=l-(1-z)P/2$.
%%(m*(1 + 2*nu)*Gamma[1 + nu]*Gamma[1 + m + nu])/Gamma[3 + m + 2*nu]
At a renormalisation scale for which $P^2$ is negligible, this yields the Mellin moments
\begin{equation}
\langle x^m\rangle_\nu = \langle x \rangle \, m \frac{ \Gamma (2 \nu +4) \Gamma (m+\nu +1)}{\Gamma (\nu +2) \Gamma (m+2 \nu +3)}\,,
\end{equation}
corresponding to the distribution $(\bar x=1-x)$
\begin{equation}
\label{phiscalarcl}
\phi_\sigma^{\rm cl}(x) = \langle x \rangle \,
\frac{2\nu+3}{2\nu+1}\frac{ (x - \bar x)^\nu \, C_1^{\nu_+}(x - \bar x)}{B(\nu+1,\nu+2)}\,,
%\frac{\Gamma (2 \nu +4)}{(2 \nu +1) \Gamma (\nu +1) \Gamma (\nu +2)}
\end{equation}
where the first moment sets the mass-scale, $C_1^{\nu_+}$ is a Gegenbauer polynomial of order $\nu_+=(\nu+1/2)$ and $B(u,v)$ is the Euler $\beta$-function.\footnote{Curiously, even for a contact-interaction, the PDA is $x$-dependent: $\phi_\sigma^{\rm cl}(x) = \langle x \rangle\,  6 (x-\bar x)$.  This contrasts with pseudoscalar mesons, for which it is $x$-independent.}

Plainly, $\phi_\sigma^{\rm cl}(x)$ is odd under $x\leftrightarrow \bar x$; and, indeed, this is a general property of the leading-twist PDA of $0^{++}$ systems, \emph{i.e}.\ owing to charge-conjugation symmetry:
\begin{equation}
\label{phiminusphi}
\phi_{0^{++}}(x) = - \phi_{0^{++}}(\bar x) \,.
\end{equation}
It follows that
\begin{equation}
\label{0ppxminusbarx}
0^{++}: \; \langle x\rangle = -\langle \bar x \rangle\,,
\end{equation}
\emph{viz}.\, on average, the valence-quark and valence-antiquark carry equal but opposite fractions of the bound-state's light-front momentum.  In addition, Eq.\,\eqref{phiminusphi} entails
\begin{equation}
\label{xeqx2}
0^{++}: \; \langle x\rangle = \langle x^2 \rangle\,,\;
\langle x^4 \rangle = 2 \langle x^3 \rangle - \langle x \rangle\,,
\end{equation}
and a countable infinity of kindred identities, each of which uniquely determines a given even moment in terms of some combination of all lower-order moments.

Eq.\,\eqref{0ppxminusbarx} contrasts starkly with the results obtained for pseudoscalar- and vector-mesons \cite{Chang:2013pq, Gao:2014bca, Ding:2015rkn, Li:2016dzv}:
\begin{equation}
0^{+-}\,,\;1^{--}: \; \langle x\rangle = +\langle \bar x \rangle\,.
\end{equation}
The analogues of Eq.\,\eqref{xeqx2} for a symmetric PDA are
\begin{equation}
\label{xeqx2sym}
0^{+-}\,,\;1^{--}:\; \langle x \rangle =\tfrac{1}{2} \langle x^0 \rangle\,, \; 4\langle x^3\rangle = 6\langle x^2 \rangle -\langle x^0\rangle\,,
\end{equation}
etc.  Pseudoscalar- and vector-mesons are considered to be $S$-wave states in the two-body quark model.  When $\langle x\rangle = \langle \bar x \rangle$, the PDA peaks at zero relative momentum, so Eq.\eqref{0ppxminusbarx} can be seen to indicate that the valence-constituents of a $0^{++}$ bound-state possess a maximal amount of relative light-front momentum.  It is now a small step to appreciate that small light-front relative momentum ensures minimal light-front angular momentum, hinting at $S$-wave primacy in pseudoscalar- and vector-mesons on the light-front, whilst maximal relative momentum promotes maximal angular momentum and hence points to an enhanced role for $P$-waves in $0^{++}$ bound-states on the light-front.

\subsection{PDAs: computational background}
In order to determine the PDAs of light-quark scalar-meson ground-states and first radial excitations, we use the direct method introduced in Ref.\,\cite{Ding:2015rkn} and exploited in Ref.\,\cite{Li:2016dzv}.  Namely, employing a symmetry-preserving truncation of the relevant gap- and Bethe-Salpeter equations \cite{Binosi:2016rxz}, we calculate the meson's Bethe-Salpeter wave function.  With that in hand, one can directly compute Mellin moments of the associated leading-twist PDA, Eq.\,\eqref{Mellinmom}, using interpolations of the numerical solutions for the propagators and Bethe-Salpeter amplitudes.  The distribution itself can be reconstructed from those moments, following the approach of Refs.\,\cite{Chang:2013pq, Cloet:2013tta}.

In order to implement this ``brute force'' approach, a factor
\begin{equation}
\label{cutofffunction}
{\mathpzc d}(k^2 r^2) = 1/(1+k^2 r^2)^{m/2}
\end{equation}
is introduced for each $\langle x^{m}\rangle$, $m\geq 1$.  The moment is then computed as a function of $r^2$, with the values subsequently fitted by a smooth function, which is used to extrapolate to $r^2=0$.  This approach typically yields reliable results for at least four nontrivial Mellin moments of each distribution considered herein.

Current-quark masses play a role in our analysis because they influence the form of the dressed-quark propagator and, consequently, meson Bethe-Salpeter amplitudes.  The propagator has the general form:
\begin{equation}
 S(p) %= \frac{1}{i \gamma\cdot p \, A(p^2,\zeta^2) + B(p^2,\zeta^2)}
= Z(p^2,\zeta^2)/[ i\gamma\cdot p + M(p^2)] \,.
\label{SgeneralN}
\end{equation}
It is obtained from a gap equation that involves a ``seed'' current-quark mass, which distinguishes the quark flavour, and is augmented by a renormalisation condition.  A mass-independent renormalisation scheme is useful, and can be implemented by making use of the scalar WGT identity and fixing all renormalisation constants in the chiral limit \cite{Chang:2008ec}.  Notably, the mass function, $M(p^2)$, is independent of the renormalisation point; and the renormalised current-quark mass is given by
\begin{equation}
\label{mzeta}
m^\zeta = Z_m(\zeta,\Lambda) \, m^{\rm bm}(\Lambda) = Z_4^{-1} Z_2\, m^{\rm bm}.
\end{equation}

Like the running coupling constant, the running mass in Eq.\,\eqref{SgeneralN} is a familiar concept; and the renormalisation group invariant (RGI) current-quark mass may be inferred via
\begin{equation}
\hat m = \lim_{p^2\to\infty} \left(\tfrac{1}{2}\ln [p^2/\Lambda^2_{\rm QCD}]\right)^{\gamma_m} M(p^2)\,,
\end{equation}
where $\gamma_m = 12/(33-2 N_f)$: $N_f$ is the number of quark flavours employed in computing the running coupling.  The chiral limit is expressed by
\begin{equation}
\label{chirallimit}
\hat m = 0\,.
\end{equation}

We work with a renormalisation scale $\zeta=\zeta_2 := 2\,$GeV and employ valence-quark RGI current masses:
\begin{equation}
\label{currentmass}
\hat m_u=\hat m_d = 5.9\,\mbox{MeV}\,,\;
\hat m_s = 137\,\mbox{MeV}\,,
\end{equation}
which correspond to one-loop evolved values:
\begin{equation}
m_{u=d}^{\zeta_2}=4.1\,\mbox{MeV}, \; m_{s}^{\zeta_2}=95\,\mbox{MeV}.
\end{equation}
The ratio $2 \hat m_s/[\hat m_u + \hat m_d] = 23$ is, perhaps, 10\% too small \cite{Agashe:2014kda}, but that is within the error typically associated with the rainbow-ladder DSE truncation, described below.

In solving the Bethe-Salpeter equation for a given meson, we adopt the Chebyshev expansion technique described as ``Method B'' in Ref.\,\cite{Maris:1997tm}, \emph{i.e}.\ each scalar function in the associated Bethe-Salpeter amplitude is expanded in terms of Chebyshev moments:
\begin{equation}
\label{Chebyshev}
^j \!F(\ell^2) := \frac{2}{\pi} \int_{-1}^{1}\!\! dx\,\sqrt{1-x^2}\,U_j(x)\,F(\ell^2,x;P^2)\,,
\end{equation}
where $\ell\cdot P = x \sqrt{\ell^2 P^2}$ and $U_j(x)$ is a Chebyshev polynomial of the second kind.  The Bethe-Salpeter equation then becomes a matrix of linear equations for the $\ell\cdot P$-independent Chebyshev moments.  The accuracy of the method depends on the number of Chebyshev moments retained for each function.  We keep as many as are necessary (typically five or six) in order to ensure stable results for all Mellin moments of a given distribution in those cases for which the signal-noise ratio in the brute-force evaluation is sufficient for the determination of a reliable value.

\begin{table}[t]
\caption{\label{tableresults}
Computed values for selected qualities of scalar and vector valence-quark-antiquark bound-states: $n=0$ labels a ground-state and $n=1$, the first radial excitation; the renormalisation scale is $\zeta_2=2\,$GeV; as appropriate, $f_\ell$ is either the leptonic decay constant in Eq.\,\eqref{fsigma} or Eq.\,\eqref{rhodecayconstantA}, and $f_\zeta$ is either the scale-dependent decay constant in Eq.\,\eqref{rhosigma} or Eq.\,\eqref{rhodecayconstantB}.  (Rainbow-ladder truncation; $m_{\cal G}=1.1\,$GeV; isospin-symmetric limit, so $u\bar u \sim d\bar d \sim u \bar u + d\bar d$; and all dimensioned quantities in GeV.)
}
\begin{tabular*}
{\hsize}
{
l|@{\extracolsep{0ptplus1fil}}
c|@{\extracolsep{0ptplus1fil}}
c|@{\extracolsep{0ptplus1fil}}
c|@{\extracolsep{0ptplus1fil}}
c|@{\extracolsep{0ptplus1fil}}
c@{\extracolsep{0ptplus1fil}}}\hline
$J^P$ &  $(q\bar q)$ & $n$ & mass & $f_\ell$ & $f_\zeta $ \\\hline
$0^+$ & $u\bar u$ & $0$ & 0.90 & 0\phantom{.028} & $\phantom{-}$0.19\phantom{7} \\
&                & $1$ & 1.47 & 0\phantom{.028} & $\phantom{-}$0.037 \\
& $u\bar s$ & $0$ & 1.08 & 0.031 & $\phantom{-}$0.18\phantom{7} \\
& $s\bar s$ & $0$ & 1.23 & 0\phantom{.028} & $\phantom{-}$0.16\phantom{7} \\\hline
$1^-$ & $u\bar u $ & 0 & 1.02  & 0.19\phantom{8} & $\phantom{-}$0.17\phantom{7} \\
          &                 & 1 & 1.24  & 0.11\phantom{8} & $-$0.046 \\\hline
\end{tabular*}
\end{table}
%% estimate accuracy of 0.19 value
% ... qbq at 2 GeV is approx 0.26^3 ... ratio rho_sigma/rho_pi at zeta=19 = (0.53/0.49)^2
% ... ratio unchanged by evolution => \rho_\sigma(2) = (0.53/0.49)^2  * 0.26^3/0.0924 - 0.222 GeV^2
% ... divide by mass => 0.25 GeV.
%% not precisely the same as Sixue's because, in RL truncation, D*w(2GeV)=(1.1)^3 is effectively smaller than D*w(19GeV)=(1.1)^3
%% ZT and Z2 ... similar?  Then value of tensor decay constant = Z2*Bo-Lin's result.
% 0.69623657*200.30 = 0.14
% 0.69623657*(-0.0419482) = -0.029
%% Lei has now done this correctly ... ZT=1.13
%% Dw=0.87^3 ... fP=153.78  fT= 141.37 ... Dw=1.1^3 ... fP=186.67  fT=226.13 ... larger, must be an artefact of enlarged Dw ... use 186.68 * 141.37/153.78 = 171.615

Since our main goal is to highlight a range of qualitative features of ground-state and radially-excited scalar and vector mesons, it is sufficient herein to employ the simplest, most widely used approximations to the gap- and Bethe-Salpeter equations, \emph{viz}.\ the rainbow-ladder (RL) truncation.\footnote{\emph{N.B}.\,Concerning ground-state PDAs, results obtained using RL truncation can be compared with those produced by the most sophisticated approximation currently available, the so-called DB kernels \cite{Chang:2011ei, Binosi:2014aea}: despite noticeable quantitative differences, they agree qualitatively in all respects \cite{Chang:2013pq, Chang:2013epa, Shi:2015esa}.}  The RL kernels are completely determined once an interaction is specified; and we use that introduced and explored in Refs.\,\cite{Qin:2011dd, Qin:2011xq}:
\begin{equation}
\label{CalGQC}
\frac{1}{Z_2^2}{\cal G}(s) = \frac{8 \pi^2}{\omega^4} D \, {\rm e}^{-s/\omega^2}
+ \frac{8 \pi^2 \gamma_m\,{\cal F}(s)}{\ln [ \tau + (1+s/\Lambda_{\rm QCD}^2)^2]} ,
\end{equation}
where $N_f=4$ in $\gamma_m$, $\Lambda_{\rm QCD}=0.234\,$GeV, $\tau={\rm e}^2-1$, and ${\cal F}(s) = \{1 - \exp(-s/[4 m_t^2])\}/s$, $m_t=0.5\,$GeV.  This interaction preserves the one-loop renormalisation-group behavior of QCD in the gap- and Bethe-Salpeter-equations \cite{Maris:1997tm}, it is consistent with modern DSE and lattice studies \cite{Boucaud:2011ug, Aguilar:2015bud}, and the infrared structure serves to ensure dynamical chiral symmetry breaking (DCSB) \cite{Horn:2016rip} and confinement, the latter through the violation of reflection positivity \cite{Stingl:1985hx, Krein:1990sf, Hawes:1993ef, Roberts:1994dr}.

Notably, as illustrated in Refs.\,\cite{Bhagwat:2006xi, Eichmann:2008ae, Qin:2011dd, Qin:2011xq}, the parameters $D$ and $\omega$ in Eq.\,\eqref{CalGQC} are not independent: with $m_{\mathpzc G}^3:=D\omega=\,$constant, one can expect numerous computed observables to be practically insensitive to $\omega$ on the domain $\omega\in[0.4,0.6]\,$GeV.  We use $\omega=0.5\,$GeV and $m_{\mathpzc G}=1.1\,$GeV, a value chosen so that corrections to RL truncation may act and draw computed results into line with empirical values \cite{Eichmann:2008ae}.  With this interaction, the RGI masses in Eq.\,\eqref{currentmass} correspond to the nonperturbatively-renormalised current-quark masses $m_{u,d}(\zeta_2)=8.5\,$MeV, $m_s(\zeta_2)=0.2\,$GeV and generate mass functions with $M_{u,d}(\zeta_2^2)=75\,$MeV, $M_s(\zeta_2^2)=253\,$MeV.\footnote{That these values are larger than usually imagined is a defect of the RL truncation.  It can be overcome by using the more complicated, realistic DCSB-improved (DB) kernels, which employ a strongly-dressed gluon-quark vertex in the gap equation and concomitant modifications of the Bethe-Salpeter kernel \cite{Chang:2011ei, Binosi:2014aea}.  As noted above, however, such improvement cannot qualitatively alter the results herein.}

Our calculated results for masses and decay constants associated with the valence-quark-antiquark core of light-quark scalar mesons are listed in Table~\ref{tableresults}.  (We define all Bethe-Salpeter amplitudes such that the zeroth Chebyshev moment of the dominant Poincar\'e-covariant is positive at large-$\ell^2$.)  Notably, the values of the scale-dependent decay constants are consistent with earlier DSE calculations \cite{Bhagwat:2006py, Qin:2011xq} and similar in size to those inferred in Refs.\,\cite{DeFazio:2001uc, Bediaga:2003zh, Yinelek:2012qda}, but, therefore, roughly a factor of two smaller than the estimates in Ref.\,\cite{Cheng:2005nb}.
%% Check Sqrt[2] factor ... (mu+md)qbq/mpi^2 = fpi^2 => fpi= 0.085 ... so Cheng use 92 MeV normalisation and their results are directly comparable to ours.
Moreover, as telegraphed following Eq.\,\eqref{CalGQC}, the masses of all listed systems are ``inflated'' by our choice $m_{\mathpzc G} = 1.1\,$GeV, leaving room for corrections to RL truncation, such as meson-meson final-state interactions, which might sometimes be considered as introducing a molecular component, to reduce the quoted mass and introduce a width \cite{Holl:2005st, Eichmann:2015cra}.

\begin{table}[t]
\caption{Mellin moments, $10^2 \langle x^m\rangle$, of the $0^+$- and $1^{--}$-meson leading-twist PDAs computed using the method described in connection with Eq.\,\eqref{cutofffunction} and Fig.\,\ref{extrap}.  The entry ``x'' indicates that the extrapolated result is not accurate to better than 5\% and is therefore discarded.  (The Bethe-Salpeter wave functions are calculated in RL-truncation with a renormalisation scale $\zeta=\zeta_2$.  Dimensioned quantities in GeV.)
\label{TableMellinMoments}
}
\begin{tabular*}
{\hsize}
{
l|@{\extracolsep{0ptplus1fil}}
c|@{\extracolsep{0ptplus1fil}}
c|@{\extracolsep{0ptplus1fil}}
c|@{\extracolsep{0ptplus1fil}}
c|@{\extracolsep{0ptplus1fil}}
c|@{\extracolsep{0ptplus1fil}}
c|@{\extracolsep{0ptplus1fil}}
c|@{\extracolsep{0ptplus1fil}}
c|@{\extracolsep{0ptplus1fil}}
c@{\extracolsep{0ptplus1fil}}}\hline
$J^P$ & $(q\bar q)$ & $n$ & $m=0$ & $1$ & $2$ & $3$ & $4$ & $5$ & $6$ \\\hline
$0^+$ & $u\bar u $ & $0$ & $\phantom{1}0\phantom{.12} $ & $\phantom{-}3.46\phantom{8}$ & $\phantom{-}3.46\phantom{8}$ & $\phantom{-}3.03\phantom{3}$ & $\phantom{-}2.59$ & $\phantom{-}2.29$ & $\phantom{-}2.10$ \\
%  0.0043756 0.0043756 -0.0038294 -0.012055 -0.017775 -0.021099
            & $u\bar u $ & $1$ & $\phantom{1}0\phantom{.12} $ & $\phantom{-}0.438$ & $\phantom{-}0.438$ & $-0.383$ & $-1.21$ & $-1.78$ & $-2.11$ \\
        & $u\bar s $ & $0$ & $\phantom{1}3.12$ & $\phantom{-}5.05\phantom{8}$ & $\phantom{-}3.99\phantom{8}$ & $\phantom{-}3.04\phantom{3}$ & x& x & x \\
% 0.040849 0.040849 0.035228 0.029488 0.024836
        & $s\bar s$ & $0$ & $\phantom{1}0\phantom{.12}$ &  $\phantom{-}4.08\phantom{8}$ & $\phantom{-}4.08\phantom{8}$ & $\phantom{-}3.52\phantom{3}$ & $\phantom{-}2.95$ & $\phantom{-}2.48$ & x \\\hline
%%
%{0.107017, 0.0535086, 0.0131463, -0.00736924}
$1^-$ & $u \bar u\,\parallel$ & 1 & $10.7$ & $\phantom{-}5.35\phantom{8}$ & $\phantom{-}1.31\phantom{8}$ & $-0.737$ & $-1.66$ & x & x \\
%-0.04572     -0.02286     -0.05150 -0.06582
          & \phantom{$u \bar u$}\,$\perp$ & 1 & $-4.57$ & $-2.29\phantom{8}$ & $-5.15\phantom{8}$ & $-6.58\phantom{3}$ & $-6.88$ & x & x \\\hline
\end{tabular*}
\end{table}

\subsection{PDAs: ground-state and first radial excitation}
One now has in hand all elements necessary for a computation of the scalar meson leading-twist PDAs via Eq.\,\eqref{Mellinmom}.  Notably, in using RL truncation the Bethe-Salpeter amplitudes we obtain describe idealised scalar bound-states, with simple valence-quark structure and zero width.  The observed scalar resonances are more complex \cite{Agashe:2014kda, Pelaez:2015qba}, but the $f_0(500)$, $K_0^\ast(800)$, $f_0(980)$, $a_0(980)$ systems do contain valence-quark-antiquark components (whose strength is currently model-dependent), and these pieces yield the leading-twist PDA.
%% [http://pdg.lbl.gov/2014/reviews/rpp2014-rev-scalar-mesons.pdf] page 10

\begin{figure}[t]
\centerline{\includegraphics[width=0.42\textwidth]{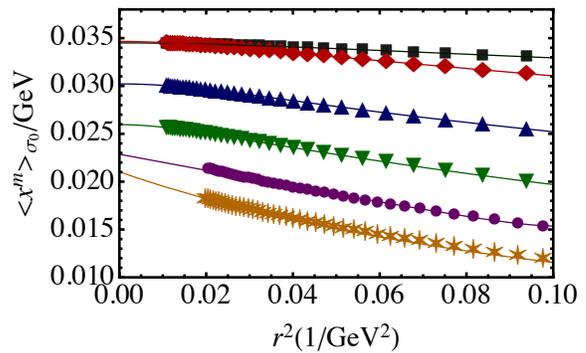}}
\caption{\label{extrap}
The $u\bar u$ moments in Table~\ref{TableMellinMoments} are the $r^2\to 0$ extrapolations of the curves depicted in this figure:
$\langle x^1\rangle$, black squares;
$\langle x^2\rangle$, red diamonds;
$\langle x^3\rangle$, blue-up-triangles;
$\langle x^4\rangle$, green down-triangles;
$\langle x^5\rangle$, purple circles;
$\langle x^6\rangle$, orange asterisks.
The curves are $[2,1]$-Pad\'e fits to the points depicted.  Other fitting forms were also employed, with no material change in the results.
}
\end{figure}

We work in the isospin symmeric limit, and the moments of the ground-state $u\bar u \sim d \bar d \sim  [u\bar u+d\bar d]$ system are listed in Table~\ref{TableMellinMoments}.  They were obtained as described in connection with Eq.\,\eqref{cutofffunction}, a procedure whose reliability is illustrated by Fig.\,\ref{extrap}.  In this case %which is the least favourable because it is the lightest system considered,
we found that sound estimates could be obtained for $m\leq 6$, although the signals for the fifth and sixth moments were lost for $r^2\lesssim 0.2\,$GeV$^2$.  Higher moments showed greater curvature and hence could not readily yield extrapolated results that were accurate to better than 5\%.  They were therefore discarded.  We verified that the same results are obtained using different forms of regulator function in Eq.\,\eqref{cutofffunction}.  It is noteworthy that the first of Eqs.\,\eqref{xeqx2} is recovered nontrivially, \emph{viz}.\ the $m=1$, $2$ moments possess different sensitivity to $r^2$, but the extrapolation curves converge to the same point; and the second is satisfied to better than $0.1\,$\%, despite all three moments having been extrapolated independently.  We used a third identity:
\begin{equation}
0^{++}: \quad \langle x^6 \rangle=3 \langle x^1 \rangle - 5 \langle x^3 \rangle+ 3 \langle x^5 \rangle\,,
\end{equation}
to aid in constraining extrapolations for the fifth and sixth moments.  %One may therefore be confident that the our implementation of the brute-force method is reliable.

Using the moments in Table~\ref{TableMellinMoments}, the PDA of a $u\bar u$ scalar bound-state may be reconstructed using the method introduced and tested in Refs.\,\cite{Chang:2013pq, Cloet:2013tta, Chang:2013epa, Segovia:2013eca, Shi:2015esa}.  We write
%%\begin{subequations}
\begin{equation}
\label{Gfitsigma}
%%\begin{align}
\phi_{\sigma}(x) = [x \bar x]^{\alpha_-} \sum_{z=0}^{z_{\rm max}} a^z_{\sigma}C_z^\alpha(x-\bar x)\,,
%%\phi_{\sigma}(x) & = m_\sigma \, \langle x^{p_\sigma} \rangle \, \varphi_\sigma(x) \,,\\
%
%%\varphi_{\sigma}(x) & = {\cal N}_\alpha\, [x \bar x]^{\alpha_-} \sum_{z=0}^{z_{\rm max}} a^z_{\sigma}C_z^\alpha(x-\bar x)\,,\\
%
%%1 & = \int_0^1 dx\, x^{p_\sigma} \varphi_{\sigma}(x)\,,
%%\end{align}
\end{equation}
%%\end{subequations}
where $\{C_z^\alpha\}$ are order-$\alpha$ Gegenbauer polynomials, $\alpha_-=\alpha-1/2$.  Notably, for $0^{++}$ systems, Eq.\,\eqref{phiminusphi} entails that the sum includes only odd Gegenbauer polynomials.
%%Consequently, $p_\sigma = 1$ in this case, whereas $p_\sigma = 0$ for flavour-unbalanced scalars.
We take $z_{\rm max}=3$; and determine the parameters $\{\alpha,a_{\sigma}^{z}\}$ via a least-squares fit that requires the odd (independent) moments of $\phi_{\sigma}(x) $ in Eq.\,\eqref{Gfitsigma} to match those in Table~\ref{TableMellinMoments}, with the results listed in Table~\ref{PDAParams}.  The associated curves reproduce the moments with a rms-relative-error of approximately 2\% (ground state) and $0.1\,$\% (excited state).

\begin{figure}[t]
\centerline{\includegraphics[width=0.42\textwidth]{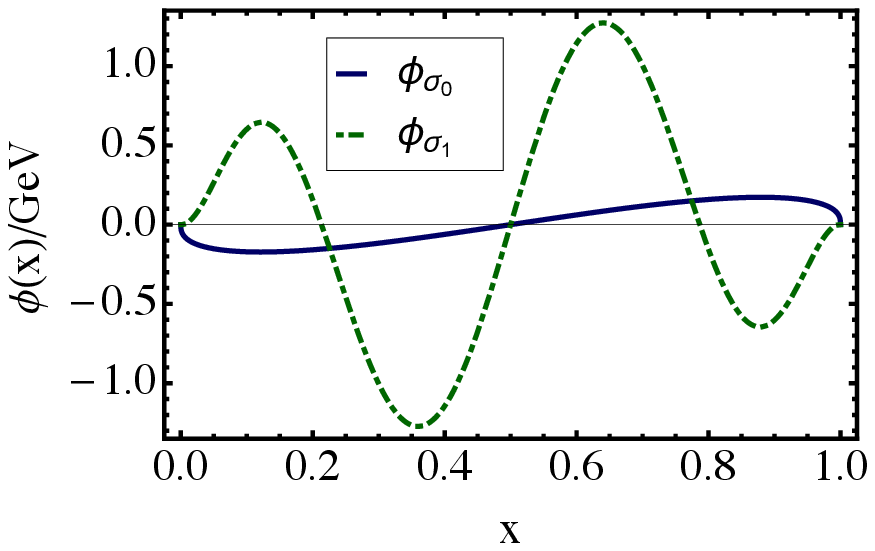}}\vspace*{2ex}

\centerline{\includegraphics[width=0.42\textwidth]{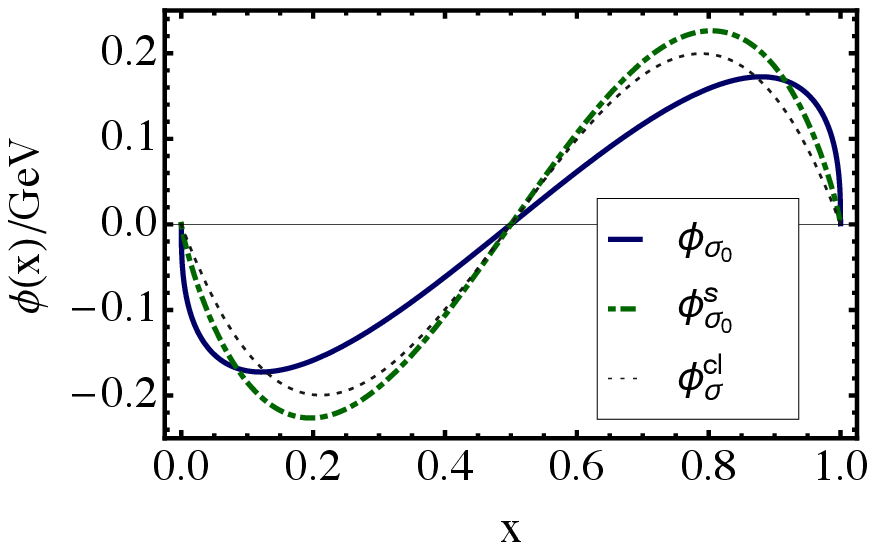}}
\caption{\label{figPDAud}
PDAs, scalar bound-states. \emph{Upper panel} --
$u\bar u$ $0^{++}$ channel: ground-state (solid blue) and first-radial excitation (dot-dashed green).
\emph{Lower panel} --
$0^{++}$ $u\bar u$  (solid blue), $s\bar s$ (dot-dashed green) channels; and dotted (black) curve, conformal-limit distribution in Eq.\,\eqref{phiscalarcl}, normalised to ensure $\langle x \rangle$ matches that of the $u\bar u$ distribution.
}
\end{figure}

The leading-twist PDAs of the ground and radially-excited $u\bar u$ $0^{++}$ bound-states specified by Eq.\,\eqref{Gfitsigma} using the parameters in Table~\ref{PDAParams} are depicted in the upper panel of Fig.\,\ref{figPDAud}.  As suggested by Eq.\,\eqref{phiscalarcl}, and in qualitative agreement with Ref.\,\cite{Cheng:2005nb}, the ground-state PDA has one zero on $0<x<1$.  On the other hand, following the pattern described in Ref.\,\cite{Li:2016dzv}, the first radial excitation has two additional zeros in this domain (three zeros altogether).  Naturally, in both cases the domains of positive and negative support are precisely balanced so that the leptonic decay constants vanish identically, with no tuning required in this symmetry-preserving calculation.\footnote{It is notable that the light-front holographic model reviewed in Ref.\,\cite{Brodsky:2014yha} yields $\phi_\sigma \equiv 0$ for $n\geq 0$, a result in conflict with both this DSE analysis and sum rules phenomenology.}

Pursuing the reasoning in Ref.\,\cite{Li:2016dzv} further, we predict that at $\zeta=\zeta_2$, the number of zeros in the leading-twist PDA of a $0^{++}$ bound state is $2n+1$, where $n$ is the radial quantum number.  However, under ERBL evolution \cite{Lepage:1979zb, Efremov:1979qk, Lepage:1980fj} there is always an $\epsilon_{\tilde n} >0 $ and a domain ${\mathpzc N}_{\;\tilde n}=\{\zeta | \zeta>0,\Lambda_{\rm QCD}/\zeta < \epsilon_{\tilde n}\}$ such that $\forall \zeta\in {\mathpzc N}_{\;\tilde n}$, the PDA for each excitation with $n<\tilde n$ has only one zero on $0<x<1$.

\begin{table}[t]
\caption{Fitted parameters that determine the leading-twist PDAs computed herein via Eq.\,\eqref{Gfitsigma} or \eqref{Gfitvector}, as appropriate.  The entry ``$0$'' in columns 5-8 indicates that the associated parameter's value is zero, owing to symmetry.  (Dimensioned quantities in GeV.)
\label{PDAParams}
}
\begin{tabular*}
{\hsize}
{
l|@{\extracolsep{0ptplus1fil}}
c|@{\extracolsep{0ptplus1fil}}
c|@{\extracolsep{0ptplus1fil}}
c|@{\extracolsep{0ptplus1fil}}
c|@{\extracolsep{0ptplus1fil}}
c|@{\extracolsep{0ptplus1fil}}
c|@{\extracolsep{0ptplus1fil}}
c@{\extracolsep{0ptplus1fil}}}\hline
$J^P$ & $(q\bar q)$ & $n$ & $\alpha$ & $a^0$ & $a^1$ & $a^2$ & $a^3$  \\\hline
%{{1, -0.0341}, {2, -0.0341}, {3, \-0.0299}, {4, -0.02568}};
$0^+$ & $u\bar u $ & 0 & $0.869$ & $0$ & $0.300$ & $0$ & $\phantom{-1}0.0$\phantom{9} \\
%
%2:60935 0.0 0.43091 0.0 -6.7801
& $u\bar u $ & 1 & $2.65\phantom{7}$ & 0 & $0.455$ & 0 & $-7.10$ \\
%
%0:88707 0.064818 0.30918 -0.17951 0.0
& $u\bar s $ & 0 & $0.887$ & $0.0648$ & $0.309$ & $-0.18$ & $\phantom{-1}0.0$\phantom{9}\\
%
%1:3571 0.0 0.6699 0.0 -0.0010
& $s\bar s $ & 0 & $1.36\phantom{7}$ & 0 & $0.671$ & 0 & $\phantom{-1}0.0$\phantom{9} \\
\hline
$1^-$
& $u\bar u$\,$\parallel$ & 1 & 1.17\phantom{7} & 1 & 0 & $-2.60$ & 0 \\
& \phantom{$u\bar u$}\,$\perp$ & 1 & 1.36\phantom{7} & 1 & 0 & $10.2$ & 0 \\ \hline
\end{tabular*}
\end{table}

Table~\ref{TableMellinMoments} also lists our computed values for the Mellin moments of the $0^{++}$ $s\bar s$ bound-state supported by the RL truncation, which are reproduced by the function in Eq.\,\eqref{Gfitsigma} when the parameters in the fourth row of Table~\ref{PDAParams} are used ($0.1\,$\% rms relative error).  The associated curve is depicted in the lower panel of Fig.\,\ref{figPDAud}, which provides a comparison between the $u\bar u$ and $s\bar s$ channels in order to illustrate the current-mass dependence of the $0^{++}$ systems.  Evidently, with increasing current-mass, both PDA extrema migrate toward $x=1/2$.
This suggests that if one considers a $0^{++}$ bound-state with mass $m_{Q\bar Q}$, then at any finite renormalisation scale there is a neighbourhood ${\mathpzc h}=\Lambda_{\rm QCD}/ m_{Q\bar Q} \simeq 0$ such that:
%approximated by the following distribution $[(1/2)^+ = (1/2)+ {\mathpzc h}]$:
%%% Expand[(((x^5 (2 x - 1)) /. {x -> (1/2) (1 - ee)}) + ((x^5 (2 x -1)) /. {x -> (1/2) (1 + ee)}))/(2 ee^2)] /. ee -> 0
%%%Expand[((((2 x - 1)^iv (2 x - 1)) /. {x -> (1/2) (1 - ee)}) + (((2 x - 1)^iv (2 x - 1)) /. {x -> (1/2) (1 + ee)}))/(2 ee^2)] /. ee -> 0
\begin{align}
\phi_{\sigma}^{Q\bar Q}(x) \propto  m_{Q\bar Q} \frac{1}{2 {\mathpzc h}}
& \left[  \delta(x-\tfrac{1}{2}^+) - \delta( \bar x - \tfrac{1}{2}^+)\right]\,,
\end{align}
where $(1/2)^+ := (1/2)+ {\mathpzc h}$, and hence $\langle x -\bar x \rangle \propto m_{Q\bar Q}$, $\langle (x -\bar x)^{(2m+1)} \rangle \approx 0$ $\forall m\geq 1$.  (Naturally, all even moments vanish.)  The growth in peak-magnitude of this PDA with $m_{Q\bar Q}$ matches that of ground-state pseudoscalar- and vector-meson PDAs, as may be inferred from Ref.\,\cite{Bhagwat:2006xi}.
% The quarks are infinitely heavy => n.kq = M and mQpQbar = 2 M.  kqz/M -> 0  ... light-front is not always the same as the infinite momentum frame ... infinite momentum frame can't be reached for static quarks.

\begin{figure}[t]
\centerline{\includegraphics[width=0.42\textwidth]{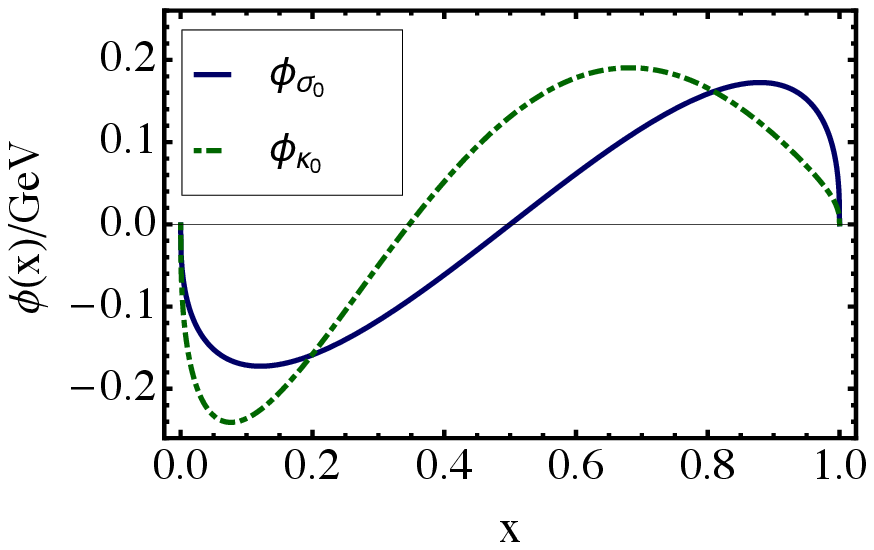}}\vspace*{2ex}

\centerline{\includegraphics[width=0.42\textwidth]{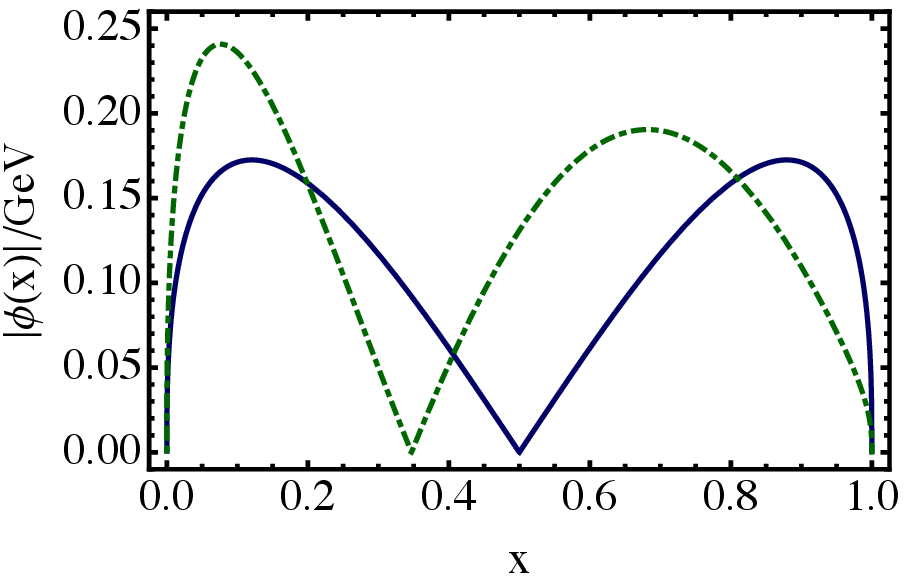}}
\caption{\label{figPDAus}
PDAs, scalar bound-states -- $0^{++}$ $u\bar u$ (solid blue) and $0^+$ $u\bar s$ (dot-dashed green): \emph{upper panel}, the PDAs themselves; and \emph{lower panel}, their absolute values.}
\end{figure}

Computed values for the moments of the $u\bar s$ bound-state are also reported in Table~\ref{TableMellinMoments}.  They are reproduced by Eq.\,\eqref{Gfitsigma} when the parameters in the third row of Table~\ref{PDAParams} are employed.  (Here, $z_{\rm max}=2$.)  Fig.\,\ref{figPDAus} compares PDAs for the $u\bar u$ and $u\bar s$ channels.  Considering the plots of $|\phi(x)|$ in the lower panel, the interior minimum of the $u\bar s$ distribution is shifted 30\% from $x=1/2$ toward $x=0$, which means that in this system the fraction of the light-front momentum carried by the $\bar s$-quark is greater than that carried by the $u$-quark.  Albeit a little larger, the magnitude of this distortion is similar to that observed in all analogous cases \cite{Braun:2004vf, ElBennich:2011py, Chen:2012txa, Shi:2015esa, Chen:2016snoPRD}.  Hence here, too, the flavour-dependence of DCSB determines the strength of $SU(3)$-flavour breaking, not the current-quark mass-difference generated by the Higgs mechanism.

On the other hand, as the heavy-light limit is approached, one expects any dependence on the heavy current-mass to factorise so that the leading-twist PDA approaches a simple, limiting form; something like, \emph{e.g}.:
\begin{equation}
\label{HLform}
\phi_{q\bar Q}(x) \sim f_{\ell}^{q\bar Q} \frac{x}{x_w^2(1-x)^3}\, {\rm e}^{-x/[x_w (1-x)]}\,,
\end{equation}
where $x_w$ is a calculable mass-independent width parameter.  (Eq.\,\eqref{HLform} is motivated by analyses in Refs.\,\cite{Grozin:1996pq, Burkardt:2001dy}.)  The images in Fig.\,\ref{figPDAus} suggest this is plausible, but the problem deserves particular attention elsewhere.
%%% http://arxiv.org/pdf/hep-ph/9607366.pdf ... leading-twist pseudoscalar PDA ~ x Exp[-x/x0] ... HQET.  This cannot become a delta-function because the wave-function of a heavy-light system is independent of the heavy-quark mass in the MQ->Infinity limit.  There is a binding energy and a characteristic size associated with the light-quark distribution.

\begin{figure}[t]
\centerline{\includegraphics[width=0.42\textwidth]{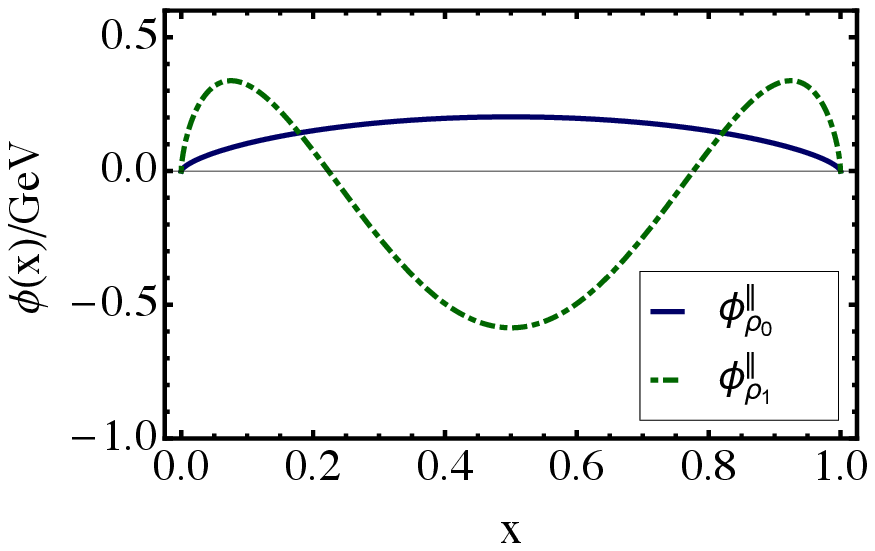}}\vspace*{2ex}

\centerline{\includegraphics[width=0.42\textwidth]{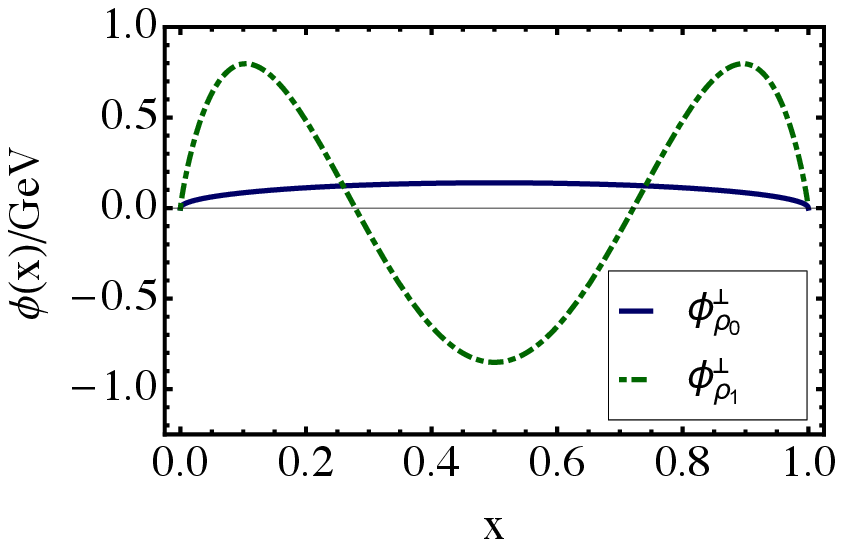}}
\caption{\label{figPDAvector1}
Vector meson PDAs: \emph{upper panel}, longitudinal polarisation; and \emph{lower panel}, transverse polarisation.  In both panels, the solid (blue) curve is the ground-state result, drawn from Ref.\,\cite{Gao:2014bca}, and the dot-dashed (green) curve is the radially excited state, computed herein.}
\end{figure}

\section{Vector mesons}
\label{SecVectors}
Within the framework detailed above, results for the leading-twist PDAs of ground-state vector mesons are reported in Ref.\,\cite{Gao:2014bca}.  Herein we compute PDAs characterising the first radial excitation of the light-quark vector meson.  Each $1^{--}$ meson possesses two independent leading-twist PDAs: $\phi_\|(x)$, $\phi_\perp(x)$, which describe, respectively, the light-front fraction of the meson's momentum carried by the quark in a longitudinally or transversely polarised system \cite{Ball:1998sk}.  The two distinct, associated decay constants are nonzero and, consequently, the related PDAs might bear some similarity to those obtainable using quantum mechanical models.  In any event, our results will serve the purpose of providing benchmarks against which other approaches can be checked, especially insofar as the consequences of symmetries are concerned.

\begin{figure}[t]
\centerline{\includegraphics[width=0.42\textwidth]{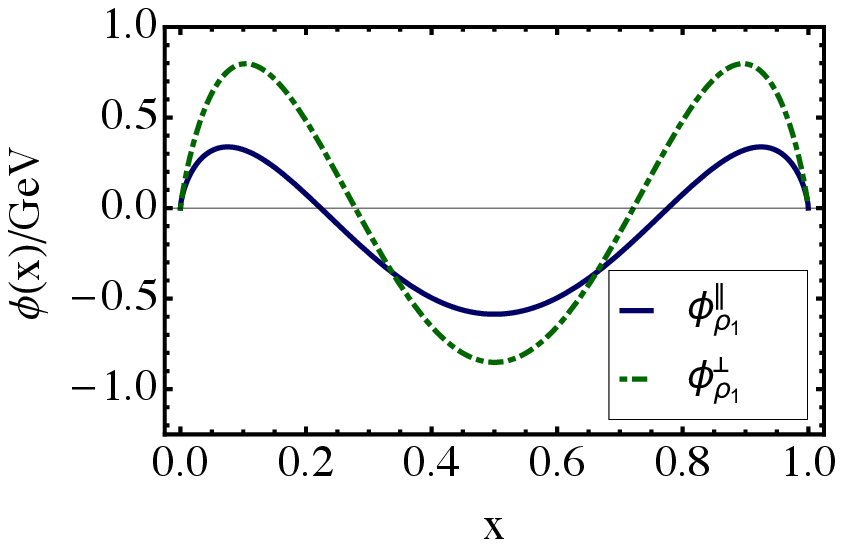}}\vspace*{2ex}

\centerline{\includegraphics[width=0.42\textwidth]{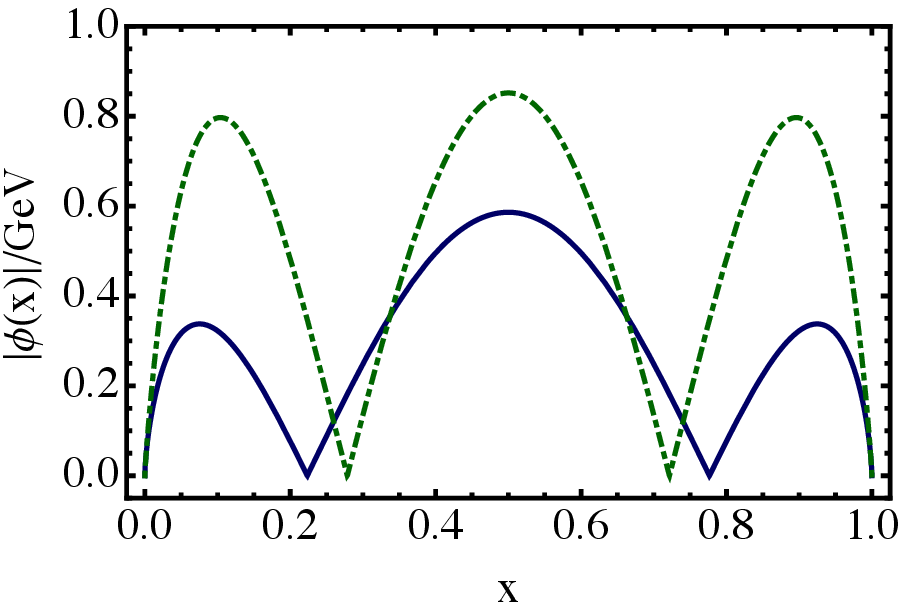}}
\caption{\label{figPDAvector2}
Vector meson PDAs: \emph{upper panel}, the PDAs themselves; and \emph{lower panel}, their absolute values.  Legend: solid (blue) curve, longitudinal projection; and dot-dashed (green) curve, transverse projection.}
\end{figure}

One must first solve the homogeneous $J^{PC}=1^{--}$ Bethe-Salpeter equation using dressed-quark propagators obtained with the light-quark mass in Eq.\,\eqref{currentmass} and the interaction in Eq.\,\eqref{CalGQC}.  (Technical details are provided elsewhere \cite{Maris:1999nt}.)  Using $m_{\cal G}=1.1\,$GeV, as in Sec.\,\ref{SecScalars}, this yields the masses in Table~\ref{tableresults}.  Having thus obtained the Bethe-Salpeter wave function, $\chi_\nu(\ell;P)$, defined as indicated by Eq.\,\eqref{chiS}, one can calculate the Mellin moments of the two independent distributions:
{\allowdisplaybreaks
\begin{subequations}
\label{momentsE}
\begin{eqnarray}
\nonumber
\lefteqn{\langle x^m \rangle_\| = \int_0^1 dx \, x^m\, \phi_\|(x)}\\
&=&
m_\rho {\rm tr}_{\rm CD} Z_2
\int_{dl}^\Lambda \frac{[n\cdot l_{q_1}]^m}{[n\cdot P]^{m+2}}\,
\gamma\cdot n \,n_\nu \chi_\nu(\ell;P)\,,\quad
\label{momentsParallel}\\
\nonumber
\lefteqn{\langle x^m \rangle_\perp = \int_0^1 dx \, x^m\, \phi_\perp(x)}\\
&=& \frac{1}{3  m_\rho^2 }  {\rm tr}_{\rm CD} Z_T
\int_{dl}^\Lambda
\frac{[n\cdot l_{q_1}]^m}{[n\cdot P]^{m}}\, \sigma_{\mu\nu} P_\mu \chi_\nu(\ell;P)\,,\quad
\label{momVT}
\end{eqnarray}
\end{subequations}
where $Z_T$ is the tensor-vertex renormalisation constant (see Appendix\,A of Ref.\,\cite{Gao:2014bca} for details), the trace is over colour and spinor indices, and our normalisation ensures:
\begin{subequations}
\label{rhodecays}
\begin{align}
\label{rhodecayconstantA}
f_\rho & = \langle x^0\rangle_\| \,,\\
\label{rhodecayconstantB}
f_\rho^\perp & = \langle x^0 \rangle_\perp\,.
\end{align}
\end{subequations}}
\hspace*{-0.5\parindent}(The factor of $1/3$ in Eq.\,\eqref{momVT} was inadvertently omitted from Eq.\,(11b) in Ref.\,\cite{Gao:2014bca}.)
The gauge-invariant quantities $f_\rho$ and $f_\rho^\perp$ are, respectively, vector-meson leptonic and tensor decay constants.  The former is a renormalisation point invariant, explaining the strength of the $\mbox{vector-meson} \to e^+ e^-$ decay, and the latter vanishes with increasing $\zeta$.  Our computed values for the ground- and radially-excited-states are listed in Table~\ref{tableresults}.
%% We renormalised the tensor vertex, with Lei finding ... at renormalization point mu=2GeV and cutoff^2=10^4 GeV^2 I get  Z_2=0.797 and Z_T=1.134 for D*omega=0.87^3;
%  Z_2=0.699 and Z_T=1.129 for D*omega=1.1^3;
%%one-loop ZT 19-> 2 = 1.16
%% Bo-Lin computed f_\| at zeta=2 and zeta=19 ... same result in both cases.
%% Different story for f_\perp.  Dw=1.1^3 ... zeta=19 GeV, f_\perp=0.14 GeV ... zeta=2, f_\perp = 0.23 > f_\|.  This is too different from ZT(2,19)^{1-loop}*fT(19) ... whereas 0.14 * fT(2)/fT(19) computed at Dw=0.87^3 is very close.
%% Plainly, Dw=1.1^3 with w=1/2 has pathalogical IR behaviour.  Probably should increase w to make things smoother.

Using the direct method described in connection with Eq.\,\eqref{cutofffunction}, we obtained results for the first five moments of $\phi_{\parallel,\perp}$.  They are listed in Table~\ref{TableMellinMoments}, and can be used to check that the identities in Eq.\,\eqref{xeqx2sym} are satisfied.

The vector meson is an eigenstate of the charge-conjugation operation with two non-vanishing decay-constants, so its PDAs take the form
\begin{equation}
\label{Gfitvector}
\phi_{\mathpzc p}(x)= f_\rho^{\mathpzc p} [x\bar x]^{\alpha_-^{\mathpzc p}}
\sum_{z=0,2,\ldots}^{z_{\rm max}} a_{\mathpzc p}^z \,C_z^{\alpha^{\mathpzc p}}(x-\bar x)\,,
\end{equation}
where ${\mathpzc p}=\parallel,\perp$ and the values of $a_{\mathpzc p}^0$ are fixed by Eqs.\,\eqref{rhodecays}.  With three even (independent) moments known accurately, we choose $z_m=2$ and determine $\{\alpha^{\mathpzc p},a_{\mathpzc p}^2\}$  via a least-squares fit that requires the moments of $\phi_{\mathpzc p}(x)$ in Eq.\,\eqref{Gfitvector} to match those in Table~\ref{TableMellinMoments}, with the results listed in Table~\ref{PDAParams}.  The associated curves reproduce the moments with a rms-relative-error of less-than 2\%.  (In this case, PDAs of almost equal quality are obtained by fixing $\alpha^{\mathpzc p}=3/2$ and fitting only $a_2^{\alpha^{\mathpzc p}}$: $a^2_{\parallel}=-2.07$, $a^2_{\perp}=9.64$.)

The computed PDAs of the vector meson's first radial excitation are plotted in Fig.\,\ref{figPDAvector1}.  At this renormalisation scale, $\zeta=\zeta_2$, each exhibits two zeros on $0<x<1$.  In fact, at $\zeta_2$, the number of zeros in the leading-twist PDA of a $1^{--}$ bound-state is $2n$, where $n$ is the radial quantum number.  The same things are true of the PDAs describing the radial excitations of pseudoscalar mesons \cite{Li:2016dzv}.  However, whereas the chiral-limit PDA of a radially excited pseudoscalar meson always has at least two zeros on $0<x<1$ \cite{Li:2016dzv}, in the case of vector mesons, even in the chiral limit, there is an $\epsilon_{\tilde n} >0 $ and a domain ${\mathpzc N}_{\;\tilde n}=\{\zeta | \zeta>0,\Lambda_{\rm QCD}/\zeta < \epsilon_{\tilde n}\}$ such that $\forall \zeta\in {\mathpzc N}_{\;\tilde n}$, the PDA for each excitation with $n<\tilde n$ is positive definite on $0<x<1$, \emph{viz}.\ the PDA is a positive, concave function, which approaches the conformal limit form ($f_\rho^{\mathpzc p} x\bar x$) under further ERBL evolution, just like the ground-state PDAs.

In Fig.\,\ref{figPDAvector2} we provide a different depiction of the vector meson PDAs.  This figure highlights that, as found for the ground-state, the PDA describing a transversely polarised vector meson exhibits greater dilation (is broader) than that of the longitudinally polarised system.  Based on Refs.\,\cite{Gao:2014bca, Ding:2015rkn}, we expect this ordering to persist with increasing current-quark mass.

\section{Conclusion}
\label{epilogue}
We calculated the leading-twist parton distribution amplitudes (PDAs) of $n=0,1$, $J^P=0^+$ mesons, where $n$ is the radial-excitation quantum number, and $n=1$, $1^{--}$ mesons, and found that, although bound-states in these channels are related by vector Ward-Green-Takahashi identities, their PDAs are very different.  In fact, associating ${\mathpzc l}=1$ with the scalar systems and ${\mathpzc l}=0$ with the vector states, then the leading-twist PDAs possess $2n+\mathpzc{l}$ zeros on $0<x<1$.  We argued that this is also true for $n\geq 2$ when the PDAs are computed at an hadronic scale.  Notably, too, the dilation characterising ground-state PDAs is also manifest in the PDAs of radial excitations.

We also considered the impact of $SU(3)$-flavour symmetry breaking.  This is a little stronger in $0^+$ systems when compared with $0^-$ states, but the size of the effect is still primarily determined by the flavour-dependence of dynamical chiral symmetry breaking, as it is in all systems studied previously.

It is worth remarking that in comparison with the leading-twist PDAs of light-quark-antiquark $0^+$ mesons computed using sum rules in Ref.\,\cite{Cheng:2005nb}, our functional forms are qualitatively in agreement.  However, we find that the decay constants reported therein are too large by a factor of two.  Our predictions for the decay constants agree with both earlier Dyson-Schwinger equation results \cite{Bhagwat:2006py, Qin:2011xq} and other estimates made using sum rules \cite{DeFazio:2001uc, Bediaga:2003zh, Yinelek:2012qda}.  It seems, therefore, that PDAs and decay constants based on those  provided herein should prove valuable in any future analyses of the nonleptonic decays of heavy mesons.

We would like to highlight that our results were obtained using the simplest symmetry-preserving truncation of the two-valence-body problem.  They will not change qualitatively with the use of a more sophisticated truncation, but it is nevertheless worth documenting the quantitative changes, and delivering predictions in future that may reasonably be considered to be definitive.

Other systems, too, deserve study.  For example, it is straightforward to show, using charge-conjugation properties, that the leptonic decay constant of $J^{PC}=1^{+-}$ mesons must vanish, just like that of scalar mesons, but the tensor decay constant need not.  On the other hand, for $J^{PC}=1^{++}$ mesons it is the other way around, \emph{i.e}.\ the tensor decay constant must vanish, but the leptonic decay constant need not.  These features will naturally constrain the behaviour of the leading-twist PDAs connected with the longitudinally and transversely polarised mesons in the associated, complementary channels to have contrasting behaviour:
\begin{equation}
\phi_\parallel^{1^{+-}} \sim \phi_\perp^{1^{++}} \sim \phi_\sigma\,, \quad
\phi_\perp^{1^{+-}} \sim \phi_\|^{1^{++}} \sim \phi_\rho\,.
\end{equation}
These expectations are qualitatively consistent with results of an existing sum rules analysis \cite{Yang:2007zt}, but they should be checked.  Light-quark tensor mesons also appear in the decays of heavy mesons, but little is known about the shape of their PDAs at the hadronic scale \cite{Cheng:2010hn}, so the techniques we described herein should also be employed in the study of tensor bound-states.

It is natural to ask whether our methods can be employed to study heavy mesons in all accessible quark-antiquark channels.  So far as attainable heavy-heavy systems are concerned, the answer is ``yes, reliably'', because the rainbow-ladder (RL) truncation is quantitatively accurate for compact states in this limit \cite{Bhagwat:2004hn, Holl:2004qn}\footnote{This does not mean that RL truncation is exact in the static limit.  Indeed, the area law \cite{Wilson:1974sk} is not simply contained in a Bethe-Salpeter kernel that can be expressed as renormalisation-group-improved one-gluon exchange.  Instead, it lies, perhaps, with  inclusion of the class of all contributions for which the so-called $H$-diagram is a generator \cite{Binosi:2016rxz}.} %
and one knows how it should sensibly be implemented \cite{Ding:2015rkn}.
%%% Caveat ... spider's web ... linear potential ... compact states = those systems that don't "feel" long range force

On the other hand, one cannot depend upon RL truncation for analyses of the arguably more interesting case of heavy-light systems.  For reasons that are well understood, connected with an imbalance between the importance of gluon-quark vertex-dressing for light-quarks as compared with heavy-quarks \cite{Bashir:2012fs}, RL truncation provides poor results for many qualities of heavy-light systems \cite{Nguyen:2009if, Rojas:2014aka}.  It is anticipated that the use of modern DCSB-improved kernels will remedy this problem with the continuum study of these important bound-states, whose decays provide a crucial window onto the Standard Model and its possible extensions.  However, such kernels are more difficult to use.  It may therefore be worth searching for a judicious modification of the RL kernels (as, \emph{e.g}.\ in Refs.\,\cite{Gomez-Rocha:2015qga, Gomez-Rocha:2016cji}) so as to obtain semi-quantitatively reliable predictions for the leading-twist PDAs of heavy-light systems in the near term.

%\section*{Acknowledgments}
\acknowledgments
We are grateful for insightful comments and suggestions from F.~Gao, S.-X.~Qin and C.~Shi.
Work supported by:
National Natural Science Foundation of China (contract nos.\ 11275097, 11475085 and 11535005);
U.S.\ Department of Energy, Office of Science, Office of Nuclear Physics, under contract no.~DE-AC02-06CH11357;
and Chinese Ministry of Education, under the \emph{International Distinguished Professor} programme.

%--
%\bibliographystyle{cj} %%%-- Correct for FBS
%\bibliography{../../../CollectedBiB}
%-
%%\bibliographystyle{../../../zProc/z10/z10KITPC/h-physrev4}
%%\bibliography{../../../CollectedBiB}

\end{document}